\documentclass[pra,aps,twocolumn,groupedaddress]{revtex4}
\usepackage{amssymb,amsmath,amsfonts,epsfig,graphicx,latexsym,bm,color}

\newcommand{\beq}{\begin{eqnarray}}
\newcommand{\eeq}{\end{eqnarray}}
\newcommand{\nn}{\nonumber}

\begin{document}
\title{Orbital superfluidity in the $P$-band of a bipartite optical square lattice}
\author{Georg Wirth, Matthias \"{O}lschl\"{a}ger, and Andreas Hemmerich}
\affiliation{Institut f\"{u}r Laser-Physik, Universit\"{a}t Hamburg, Luruper Chaussee 149, 22761 Hamburg, Germany}
\date{\today}
\maketitle
\date{\today}
\textbf{The successful emulation of the Hubbard model \cite{Fis:89} in optical lattices \cite{Jak:98, Gre:02} has stimulated world wide efforts to extend their scope to also capture more complex, incompletely understood scenarios of many-body physics \cite{Lew:07,Blo:08}. Unfortunately, for bosons, Feynmans fundamental "no-node" theorem \cite{Fey:72} under very general circumstances predicts a positive definite ground state wave function with limited relevance for many-body systems of interest. A promising way around Feynmans statement is to consider higher bands in optical lattices with more than one dimension \cite{Wu:09}, where the orbital degree of freedom with its intrinsic anisotropy due to multiple orbital orientations gives rise to a structural diversity, highly relevant, for example, in the area of strongly correlated electronic matter. In homogeneous two-dimensional optical lattices, lifetimes of excited bands on the order of a hundred milliseconds are possible \cite{Isa:05,Liu:06} but the tunneling dynamics appears not to support cross-dimensional coherence \cite{Isa:05, Mue:07}. Here we report the first observation of a superfluid in the $P$-band of a bipartite optical square lattice with $S$-orbits and $P$-orbits arranged in a chequerboard pattern. This permits us to establish full cross-dimensional coherence with a life-time of several ten milliseconds. Depending on a small adjustable anisotropy of the lattice, we can realize real-valued striped superfluid order parameters with different orientations $P_x \pm P_y$ or a complex-valued $P_x \pm i P_y$ order parameter, which breaks time reversal symmetry and resembles the $\pi$-flux model proposed in the context of high temperature superconductors \cite{Mar:89}. Our experiment opens up the realms of orbital superfluids to investigations with optical lattice models.}

Orbital optical lattices permit to prepare exciting new many-body scenarios with relevant counterparts in the area of strongly correlated electronic matter, where the orbital degree of freedom with its rich structure of possible orientations gives rise to a wealth of interesting phenomena. For example, in the transition-metal oxides orbital physics is believed to be a key element for understanding their metal-insulator transitions, superconductivity, or colossal magnetoresistance. Recent theoretical proposals have considered, for example, the formation of multiflavor and multiorbital systems \cite{Isa:05,Liu:06,Kuk:06, Xu:07}, supersolid quantum phases in cubic lattices \cite{Sca:05,Sca:06}, quantum stripe ordering in triangular lattices\cite{Wu:06}, or Wigner crystallization in honeycomb lattices \cite{Wu:07}. The recent prediction, that the life time of atoms in higher Bloch bands can be unexpectedly long, has further strengthened the potential significance of orbital physics in optical lattices \cite{Isa:05,Liu:06}. Previous experiments have demonstrated that the excitation of higher Bloch bands is in fact possible \cite{Bro:05, Koe:05}. In a recent experiment, investigating a homogeneous quasi one-dimensional (1D) lattice of bosons, transient partial coherence in the $P$-band has been observed \cite{Mue:07}. In an extension to a two-dimensional square lattice, however, no cross-coherence could be established, which had previously been predicted in Ref. \cite{Isa:05} as a consequence of the small transverse tunneling rate of the $P$-orbitals. This finding raises the question whether the implementation of non-trivial coherent orbital physics in the $P$-band, involving the freedom of orbital orientation, is at all possible in a homogeneous lattice potential. 

In this article we show that orbital superfluidity can be accomplished by an appropriate management of the tunneling junctions in a bipartite square optical lattice. We report the first observation of a 2D orbital superfluid in the $P$-band. A key feature responsible for the formation of full cross-dimensional coherence is the composition of the lattice of two types of sites with different well depths arranged as the black and white domains of a chequerboard. While the deeper wells provide local $P$-orbits, which mainly support longitudinal tunneling, the shallow wells contribute isotropic $S$-orbits, which act as cross-dimensional tunneling links. By means of a population swapping technique we can selectively excite a large fraction of the atoms into the lowest $P$-band. While initially the entire second Brillouin zone is smoothly populated and the sample does not show any coherence, after a few milliseconds a long-lived $P$-band superfluid arises, which exhibits full coherence in both lattice dimensions. Only after times on the order of several ten ms coherence is lost as the atoms undergo collisions repopulating the ground state $S$-band. Depending on the adjustment of a small anisotropy in our lattice potential, which breaks the four-fold rotational symmetry of the $P$-band, we observe striped superfluid orders with different orientations or, if isotropy of the $P$-band is established, a genuinely  complex-valued superfluid order, which exhibits staggered orbital currents. This latter case breaks time reversal symmetry, indicated by the presence of a sheet of staggered vortices with a unit flux quantum per plaquette. It realizes a variant of the $\pi$-flux model \cite{Mar:89} proposed in connection with cuprate superconductors and represents a striking example beyond Feynmans no-node theorem.

\begin{figure}
\includegraphics[scale=.35, angle=0, origin=c]{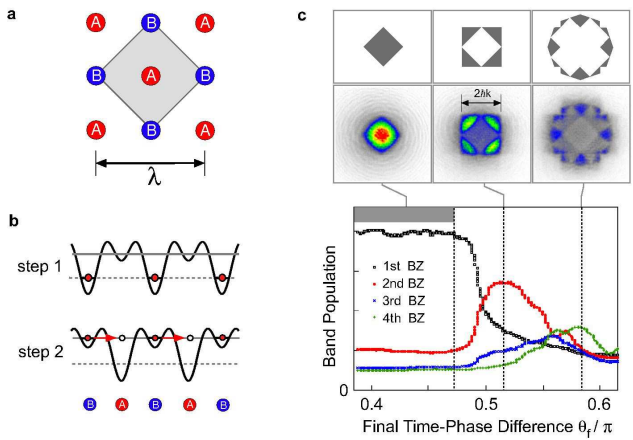}
\caption{\label{Fig.1} \textbf{Population of excited bands.} \textbf{a}, The lattice comprises two classes of lattice sites denoted by $\mathcal{A}$ and $\mathcal{B}$. The grey area denotes the Wigner Seitz unit cell of the $\mathcal{A}$-sublattice. \textbf{b}, Experimental sequence used to populate excited bands. \textbf{c}, The populations within the first four Brillouin zones (BZ) are plotted versus the final value $\theta_f$ adjusted in step 2 of the sequence in \textbf{b}. The insets show illustrations of the first, second and forth Brillouin zones (upper row) and their observed populations for values of $\theta_f$ within the grey bar and at the dashed lines (lower row).}
\end{figure}

The core of our experimental set-up is a (quasi 2D) square optical lattice, composed of two classes of (tube-shaped) lattice sites (denoted as $\mathcal{A}$ and $\mathcal{B}$) arranged as shown in Fig.1(a). The accordant optical potential is formed by crossing two optical standing waves oriented along the $x$- and $y$-axes with linear polarizations parallel to the $z$-axis, produced in the two branches of a Michelson interferometer (details are deferred to the methods section). This allows us to adjust the time phase difference between the standing waves $\theta$ to any desired value \cite{Hem:92}. For $\theta < \pi/2$ the $\mathcal{A}$-sites are more shallow than the $\mathcal{B}$-sites and vice versa.

Two different kinds of anisotropies can occur in the optical potential. The first type, which affects the structure of the energy minima in the $P$-band and the corresponding superfluid order, results from the fact that due to imperfect reflection optics the intensity ratios between the two traveling waves forming each of the two standing waves are generally unequal. In order to control this anisotropy, appropriate polarization optics in one of the standing waves lets us rotate the linear polarization of the incident beam out of the $z$-axis by a small angle $\alpha$ but maintain the polarization of the reflected beam parallel to the $z$-axis. Tuning of $\alpha$ allows us to adjust the structure of the energy minima in the $P$-band, which displays a four-fold rotational symmetry if $\alpha = \alpha_{\textrm{iso}} \approx\pi /5$. For $\alpha < \alpha_{\textrm{iso}}$ or $\alpha > \alpha_{\textrm{iso}}$ this symmetry is broken. A second kind of anisotropy results from an imbalance of the intensities in both interferometer axes, which locally is unavoidable due to the use of finite sized Gaussian beams. If $\alpha \approx \alpha_{\textrm{iso}}$, this anisotropy does not notably affect the structure of the $P$-band energy minima, where condensation takes place, and thus it is irrelevant for the structure of the superfluid order observed in our experiment. For details we refer to the methods section.

We may populate excited bands via the following two-step population swapping procedure related with techniques used in Ref. \cite{And:07} (cf. Fig.1(b)). Initially a Bose-Einstein condensate (BEC) of Rubidium ($^{87}$Rb) atoms is prepared and the lattice potential is activated with $\theta = \theta_i \approx 0.38\,\pi$, such that the well depth of the $\mathcal{B}$-sites is about 4.5 times larger than that of the $\mathcal{A}$-sites (cf. methods section). A ground state lattice is thus formed with the shallow $\mathcal{A}$-wells practically not contributing to the tunneling dynamics. In the second step, $\theta$ is rapidly increased (within 0.2 ms, which is shorter than the nearest neighbor tunneling time) to a final value $\theta = \theta_f > \pi/2$ such that now the $\mathcal{A}$-wells are significantly deeper than the $\mathcal{B}$-wells (cf. Fig.1(b)). Depending on the chosen value of $\theta_f$, population is transferred to excited bands according to the graphs shown in Fig.1(c). Expectedly excitations to higher bands are not observed for $\theta_f < \pi/2$. In this domain the $\theta$-independent background population in the excited bands is a consequence of imperfections in our adiabatic band mapping procedure and a background of uncondensed atoms arising outside the image plane because of a technical pecularity of our BEC production \cite{Kli:10}. For suitable values of $\theta_f > \pi/2$ the second (lower $P$-band) and the forth (lowest $D$-band) become the most populated bands. Notably, the third band (upper $P$-band) population always remains small.

\begin{figure}
\includegraphics[scale=0.3, angle=0, origin=c]{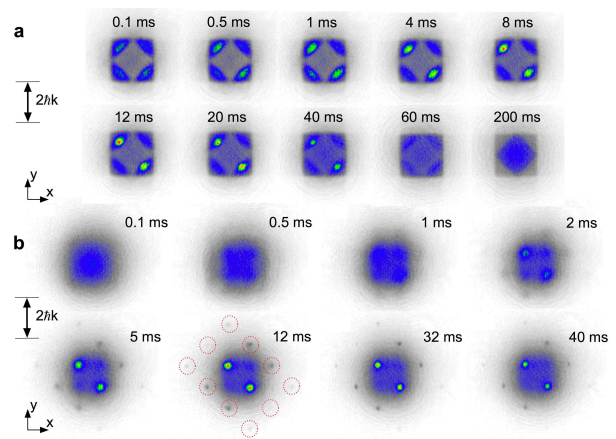}
\caption{\label{Fig.2} \textbf{Evolution of band populations and momentum distribution.} \textbf{a}, Mapping of band populations for increasing hold times in the lattice. \textbf{b}, Corresponding momentum spectra. The red dashed circles in the 12 ms spectrum highlight the higher order Bragg peaks. The lattice axes are oriented in the $x$ and $y$-directions. The case $\alpha = 0$ is shown.}
\end{figure}

For the observations reported here, the population ratio between the second and the first band is maximized ($\theta_f = 0.53\,\pi$). After a variable hold time, we map the populations in the different Brillouin zones or record the momentum distribution of the atoms. We first studied the case $\alpha = 0$, when the four-fold rotational symmetry of the $P$-band is broken. In Fig.2(a) the distribution of the atomic population among the different Brillouin zones is shown for hold times between 0.1 ms and 200 ms. After 0.1 ms the second Brillouin zone is nearly uniformly populated. As time evolves, sharp peaks arise at the two finite quasi-momenta $\textbf{K}_{(-,+)}$ and $\textbf{K}_{(+,-)}$, developing maximal contrast at hold times around 12 ms and persisting for times up to 40 ms. Here, $\textbf{K}_{(\pm,\pm)} \equiv \frac{1}{2}(\pm \hbar k, \pm \hbar k)$ with $k\equiv 2\pi/\lambda$ and $\lambda$ denotes the wavelength of the lattice beams. This indicates that a collision aided condensation process occurs. The momenta $\textbf{K}_{(-,+)}$ and $\textbf{K}_{(+,-)}$ are located on the boundary between the first and the second Brillouin zone. As the hold time is increased beyond a few ten milliseconds the sharp peaks disappear and the first Brillouin zone is repopulated, which is most clearly seen for long hold times above hundred milliseconds. In the corresponding momentum spectra (cf. Fig.2(b)) past 2 ms hold time pronounced maxima at the momenta $\textbf{K}_{(-,+)}$ and $\textbf{K}_{(+,-)}$ arise accompanied by higher order Bragg peaks (highlighted for the 12 ms picture by dashed red circles in Fig.2(b)), which indicates the emergence of a finite momentum superfluid. The absence of a zero momentum peak clearly discriminates this superfluid from the conventional superfluid observed in ground state lattices. When we carefully tune $\alpha$ in order to adjust isotropy of the $P$-band (the value of $\alpha$ has to match $\alpha_{\textrm{iso}}$ to better than $\delta\alpha \approx \pm \pi/100$ (cf. methods section)) a different superfluid order emerges. As shown in Fig.3(a) we then observe four condensation momenta $\textbf{K}_{(-,+)}$, $\textbf{K}_{(+,-)}$, $\textbf{K}_{(+,+)}$, and $\textbf{K}_{(-,-)}$ in the corresponding momentum spectra (and similarly in the Brillouin zone mappings). As is shown in Fig.3(b) we may invert the initial anisotropy adjusting $\alpha >\alpha_{\textrm{iso}} + \delta\alpha$ and recover condensation at two momenta, however with an orientation rotated by $\pi/2$ with respect to the observations for $\alpha <\alpha_{\textrm{iso}} - \delta\alpha$. Comparable condensation time-scales and condensate fractions are found independent of the value of $\alpha$.

\begin{figure}
\includegraphics[scale=.38, angle=0, origin=c]{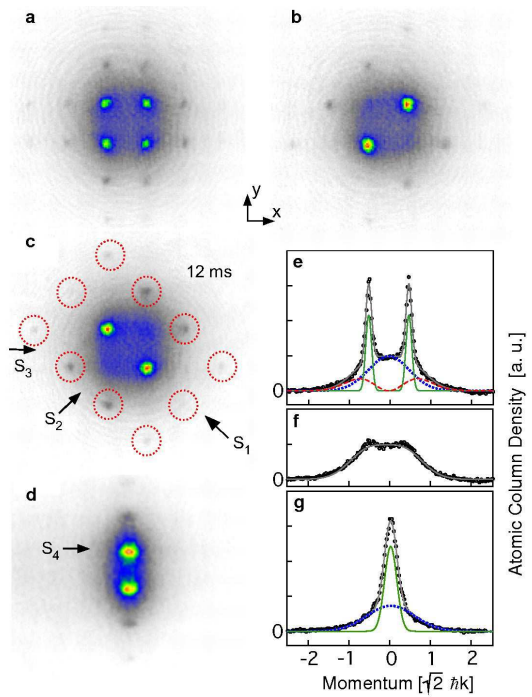}
\caption{\label{Fig.3} \textbf{Tuning of anisotropy, condensate fractions.} \textbf{a}, Momentum spectrum for an isotropic $P$-band $\alpha = \alpha_{\textrm{iso}}$. \textbf{b}, momentum spectrum for the case of $\alpha > \alpha_{\textrm{iso}}$. \textbf{c}, The 12 ms graph of Fig.2(b) is replotted. \textbf{d}, Side view of the momentum spectrum in \textbf{c} along the direction $\textrm{S}_3$. \textbf{e}, Sectional view along the direction $\textrm{S}_1$ in \textbf{c}. \textbf{f}, Sectional view along the direction $\textrm{S}_2$ in \textbf{c}. \textbf{g}, Sectional view along the direction $\textrm{S}_4$ in \textbf{c}.}
\end{figure}

In Fig.3(c-g) the 12 ms momentum spectrum of Fig.2(b) (replotted in Fig.3(c)) is analyzed more quantitatively. In Fig.3(d) a side view of this spectrum along the direction denoted by $\textrm{S}_3$ in Fig.3(c) is shown. In Fig.3(e) a sectional view along the direction denoted by $\textrm{S}_1$ in Fig.3(c) is shown. The solid grey line, matching well with the observed data (black circles), shows a fit to a sum of the (tight binding) momentum Wannier-functions of the $S$- and $P$-bands, to account for the incoherent fractions in both bands, and of two equal Gauss-functions separated by $\sqrt{2} \hbar k$, which account for the condensate fraction. The widths of the $S$-band Wannier-function and the double Gauss-function as well as the individual amplitudes of the three functions are used as fit parameters. The contributions of the $P$- and $S$-band Wannier-functions and the double Gauss-function are shown by the dashed red, dotted blue and solid green curves, respectively. The sectional view in Fig.3(f) is along the direction denoted by $\textrm{S}_2$ in Fig.3(c). The grey line repeats the incoherent fraction obtained from the data in Fig.3(e). In Fig.3(g) a section along the direction denoted $\textrm{S}_4$ in Fig.3(c) is shown. Two Gauss-functions are used to fit the data yielding the coherent and incoherent fractions given by the solid green and dotted blue curves. Figs.3(e-g) clearly show that a substantial part of the atoms is condensed. 

\begin{figure}
\includegraphics[scale=.38, angle=0, origin=c]{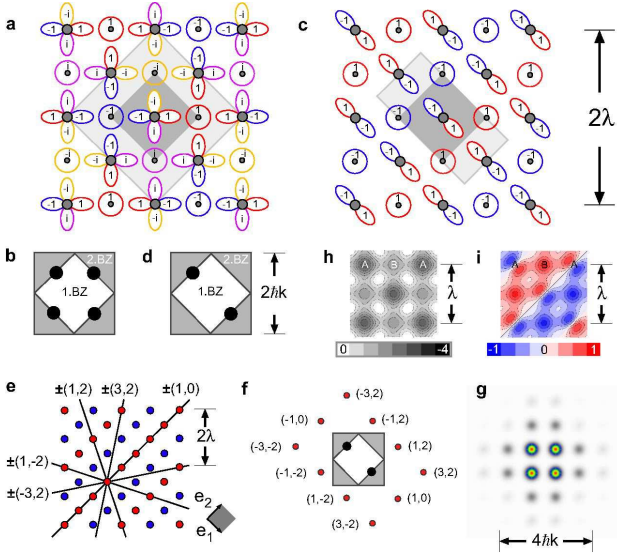}
\caption{\label{Fig.4} \textbf{Nature of superfluid order.} \textbf{a}, Orbital ordering within the $P$-band for $\alpha =\alpha_{\textrm{iso}}$ . \textbf{b}, Quasi-momentum representation corresponding to \textbf{a} (BZ = Brillouin zone). The four black disks denote the contributing quasi-momenta. \textbf{c}, Orbital alignment, if the lattice potential exhibits a small anisotropy ($\alpha < \alpha_{\textrm{iso}}$).  \textbf{d},  Quasi-momentum representation corresponding to \textbf{c}. The dark (light) grey areas in \textbf{a} and \textbf{c} show the unit cell of the lattice potential (order parameters). \textbf{e}, Schematic representation of the phase ordering in \textbf{c}. The black lines show the orientation of plane wave fronts (subscripted by their Miller indices with respect to the primitive vectors $\textbf{e}_1,\textbf{e}_2$) compatible with the imposed phase order. \textbf{f}, Positions of the Bragg resonances (small red disks) in momentum space corresponding to the wave fronts in \textbf{e}. \textbf{g}, momentum spectrum calculated from the Bloch function corresponding to \textbf{a} and \textbf{b}. \textbf{h}, Lattice potential for $\theta = 0.56\,\pi$, $\alpha  = 0$ and a well depth parameter $\bar V_0 = 6.2 \, E_{\textrm{rec}}$. \textbf{i}, Bloch-function $\phi_{\textbf{K}_{(+,-)}}(x,y)$ for the potential in \textbf{h} corresponding to the momentum components $\textbf{K}_{(+,-)}$ and $\textbf{K}_{(-,+)}$ in \textbf{d}.}
\end{figure}

The nature of the observed superfluid order parameters can be understood extending arguments discussed in Refs.\cite{Isa:05,Liu:06}. In the tight binding limit the wave function near a lattice site resembles a harmonic oscillator state. Since the atoms reside in the $P$-band, at the deeper $\mathcal{A}$-sites the possible states are superpositions of local $P_x$ and $P_y$ orbitals, which in absence of collisional interactions are energetically degenerate, if $\alpha =\alpha_{\textrm{iso}}$. The repulsive interaction favors a maximum angular momentum at these sites with an orbit configuration $P_x \pm i P_y$ corresponding to angular momentum quanta $\pm \hbar$ per atom \cite{Isa:05,Liu:06}. Since the population swapping sequence ends with the lowest energy $S$-state of the shallow $\mathcal{B}$-wells tuned into resonance with the $P$-state of the deeper $\mathcal{A}$-wells, an approximate local $S$-orbit is to be expected near the $\mathcal{B}$-sites. In order to maximize intersite hopping, the local phases of adjacent orbits should match at their tunneling junction. This suggests that the ground state within the $P$-band is characterized in configuration space by the complex order parameter illustrated in Fig.4(a). A $2\lambda \times 2\lambda$ section of the lattice is shown with the deep $\mathcal{A}$-sites and the shallow $\mathcal{B}$-sites indicated by large and small grey disks. The plotted numbers denote the local phases of the wave function. The superposition of Bloch functions $\phi_{\textbf{K}_{(+,-)}}(x,y) + i\, \phi_{\textbf{K}_{(+,+)}}(x,y)$ for the two inequivalent quasi-momenta $\textbf{K}_{(+,-)}$ and $\textbf{K}_{(+,+)}$ (derived from a band calculation for the potential in Eq. (1) of the methods section including the lowest nine bands for $\alpha =\alpha_{\textrm{iso}}$ and $\theta = 0.53\,\pi$) reproduces this orbital order. The two pairs of equivalent quasi-momenta $\textbf{K}_{(\pm,\pm)}$ and $\textbf{K}_{(\mp,\pm)}$ (plotted in (b) with respect to the first and second Brillouin zones) represent the four degenerate energy minima of the lowest $P$-band, where the system is expected to condense (cf. Fig.5 of methods section). The order in Fig.4(a) has remarkable properties. With respect to the $\mathcal{A}$-sites it exhibits a pattern of staggered orbital angular momenta. With respect to the $\mathcal{B}$-sites it exhibits vortical plaquette currents and thus resembles the $\pi$-flux phase discussed in the context of high $T_c$-superconductors \cite{Mar:89} and the staggered vortex superfluid, recently discussed in Refs. \cite{Hem:07, Lim:08, Lim:10}.

If $\alpha$ is adjusted below $\alpha_{\textrm{iso}}$, a difference in the energies of the $P_x \pm P_y$-orbits is introduced, which may exceed the collisional energy gain of the maximum angular momentum orbits. A band calculation for the case $\alpha = 0$ (and $\theta = 0.53\,\pi$) confirms that the energies of the band minima at $\textbf{K}_{(+,-)}$  and $\textbf{K}_{(-,+)}$ are lowered with respect to those at $\textbf{K}_{(+,+)}$  and $\textbf{K}_{(-,-)}$ by a few thousandth of the atomic recoil energy $E_{\textrm{rec}} \equiv \hbar^2k^2/2m$ ($m$ denotes the atomic mass) corresponding to a few percent of the energy width of the band (cf. methods section). The system then condenses into the striped state sketched in Fig.4(c), which exhibits zero local angular momentum. Instead of local $(P_x \pm i P_y)$-orbits $(P_x - P_y)$-orbits arise at the $\mathcal{A}$-sites. The condensation momenta $\textbf{K}_{(+,-)}$  and $\textbf{K}_{(-,+)}$ are shown in Fig.4(d). An equivalent state composed of the quasi-momenta $\textbf{K}_{(+,+)}$  and $\textbf{K}_{(-,-)}$ (corresponding to local $(P_x + P_y)$-orbits) arises, if $\alpha > \alpha_{\textrm{iso}}$. Remarkably, both order parameters in Fig.4(a) and (c) break the translational symmetry of the lattice potential, as is seen from their different unit cells plotted in the figure. 

The order parameters shown in Fig.4(a) and (c) readily explain the higher order Bragg peaks observed in the corresponding momentum spectra. In Fig.4(e) the $\mathcal{B}$-sublattice is replotted for the case of the striped superfluid order (Fig.4(c)) with the red and blue discs denoting sites with opposite local phase. The black lines show the wave fronts of possible plane waves that can propagate through the lattice without suffering destructive interference due to the imposed phase order. These planes, which are the lattice planes comprising only $\mathcal{B}$-sites with the same local phase, are denoted by their Miller indices. In Fig.4(f) the expected Bragg resonances in momentum space are plotted in agreement with the observations in Fig.2(b). Similar arguments apply for the case of the staggered angular momentum order in Fig.4(a). For this case we show in Fig.4(g) the momentum spectrum calculated by Fourier transforming the Bloch-function corresponding to the condensation momenta in Fig.4(b). 

For the anisotropic lattice the observed momentum spectrum of Fig.3(c) uniquely determines the order parameter of Fig.4(c). The isotropic case, however, requires a more careful discussion. The coexistence of regions with real-valued striped order but different orientations could in principle lead to a similar momentum spectrum as that of Fig.3(a). Note, however, that an uncontrolled local anisotropy, which could lead to fragmentation into striped phases, is not available in our experiment and would not be compatible with the extreme sensitivity regarding the adjustment of $\alpha$ required to observe the spectrum in Fig.3 (a). Considerations similar to those discussed, for example, in Ref. \cite{Mue:06} suggest that for globally degenerate condensation points, as realized in our experiment for $\alpha = \alpha_{\textrm{iso}}$, a coexistence scenario with its excess tunneling energy at the phase boundaries is energetically unfavorable for small repulsive interactions, although a stringent theoretical proof is yet an open problem. We thus may claim high plausibility for the conclusion that the momentum spectrum observed in Fig.3(a) in fact originates from the order parameter of Fig.4(a). A unique proof, however, remains to be an exciting challenge for future theoretical and experimental efforts.

It is instructive to observe the important role of the local $S$-orbits in the shallow $\mathcal{B}$-wells. For simplicity, we consider the striped state. As is seen from Fig.4(c) the $S$-orbits are essential for establishing coherence perpendicular to the direction of alignment of the $(P_x - P_y)$-orbits. Transverse tunneling between $P$-orbits is typically an order of magnitude weaker than axial tunneling \cite{Isa:05, Liu:06}. Tranverse tunneling between $\mathcal{A}$-wells is thus expected to occur predominantly via the $\mathcal{B}$-wells. This picture is well confirmed, if the actual lattice potential (cf. Fig.4(h)) is used to evaluate the Bloch-function (cf. Fig.4(i)) corresponding to the condensation momenta in Fig.4(d). For better visibility of the P-orbits we have used $\theta = 0.56\,\pi$ here. An analog consideration applies for the staggered angular momentum order in Fig.4(a). In fact, if resonant tunneling between $\mathcal{A}$- and $\mathcal{B}$-wells is suppressed by adiabatically switching off the $\mathcal{B}$-wells after the superfluid has been established, a rapid loss of transverse coherence is observed in the momentum spectra. Furthermore, a significant reduction of the lifetime of atoms in the $P$-band is observed. The latter finding is understood by noting that collisions within the $\mathcal{B}$-wells, where the atoms are in their local ground state, cannot reduce the $P$-band population. If, however, the atoms are constrained to reside exclusively in the $\mathcal{A}$-wells, enhanced collisional decay to the lowest band is expected. Long $P$-band life times in a bipartite lattice have been previously predicted in Ref. \cite{Sto:08}.  

In summary, we have reported the observation of orbital superfluidity in the $P$-band of a bipartite optical square lattice, which exhibits full cross-dimensional coherence with a life-time of several ten milliseconds. Our observations are explained by finite momentum superfluid order parameters comprising $S$-orbits and aligned $P$-orbits arranged in a pattern of staggered local phases. If the $P$-band is adjusted to be isotropic, the emerging complex-valued superfluid order breaks time reversal symmetry, indicated by vortical plaquette currents. Our experiment paves the way for exploring unconventional superfluids in optical lattices with potential relevance for material systems. We note that preliminary observations also show superfluidity with complex-valued order in the $D$- and $F$-bands.  

\section*{Methods Summary}
\begin{figure}
\includegraphics[scale=.32, angle=0, origin=c]{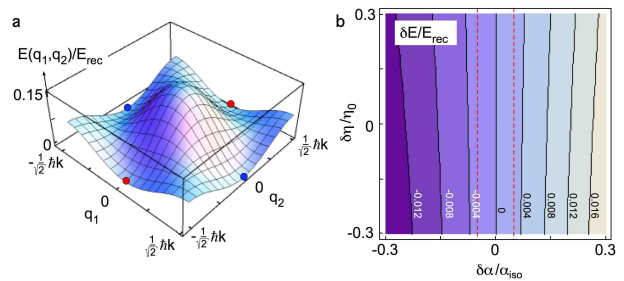}
\caption{\label{Fig.5} \textbf{Anisotropy and $P$-band energy minima.} \textbf{a}, Energy surface $E(q_1,q_2)$ of $P$-band plotted within the first Brillouin zone for $\alpha = \alpha_{\textrm{iso}}$. The red and blue discs indicate the local minima. Discs indicating equivalent minima share the same color. \textbf{b}, The energy difference $\delta E$ of two inequivalent local energy minima is plotted versus $\delta \alpha$ and $\delta \eta$.}
\end{figure}
\textit{Optical Potential} We couple a laser beam at a wavelength $\lambda=1064\,$ nm and $1/e^2$ radius $w_0 = 100\,\mu$m to a Michelson interferometer with its two branches crossed at a right angle in order to obtain the optical potential
\beq
\label{M.1}
\nn
- \frac{\bar V_0}{4} \, e^{-\frac{2z^2}{w_0^2}} | \, \eta \,\left[  \left(\hat z \,\cos{(\alpha)}  + \hat y \,\sin{(\alpha)} \right)\,e^{i k x}  + \epsilon \, \hat z \,e^{-i k x} \right] \\
+ \, e^{i \theta} \hat z \, \left(e^{i k y} + \epsilon \, e^{-i k y} \right) |^2 \,,\qquad
\eeq
where $\eta$ accounts for a small difference in the powers directed to both interferometer branches (a typical value for $\eta$ is $\eta_0 \approx 0.95$), $\epsilon \approx 0.81$ accounts for the power reduction in the retro-reflected beams due to imperfect optics, and the angle $\alpha$ permits us to adjust the degree of anisotropy introduced if $\epsilon \neq 1$. $\hat x, \hat y, \hat z$ are the unit vectors in the respective directions. An isotropic $P$-band arises if $\cos{(\alpha)} \approx \epsilon$, which amounts to $\alpha = \alpha_{\textrm{iso}} \approx \pi/5$. The four local energy minima arising in the $P$-band at the edge of the first Brillouin zone (cf. Fig.5(a)) are then degenerate. Tuning of $\alpha$ and $\eta$ affects the energy minima of the $P$-band in a notedly different way. Remarkably, if $\alpha = \alpha_{\textrm{iso}}$ is adjusted, the structure of the $P$-band energy minima is not at all modified by tuning $\eta$. This is shown in Fig. 5(b), where the energy difference $\delta E$ of the two inequivalent local energy minima of the $P$-band (red and blue discs in (a)) is plotted versus $\delta \alpha \equiv (\alpha-\alpha_{\textrm{iso}})/\alpha_{\textrm{iso}}$ and $\delta \eta \equiv (\eta- \eta_0) / \eta_0$. Within the dashed (red) lines we observe the momentum spectrum of Fig.3(a) while on the left and right the spectra of Fig.3(b) or (c) are found. The insensitivity of the $P$-band minima against changes of $\eta$ for $\delta \alpha = 0$ is a highly relevant feature, since locally an imbalance of the standing wave intensities is unavoidable due to the use of finite sized Gaussian beams. The time-phase difference $\theta$ is proportional to the path length difference of the two interferometer branches, which is actively stabilized with the help of a weak additional laser beam to better then $\pi/200$. A change of $\theta$ by $\pi/4$ can be obtained in less than 0.2 ms.

\textit{Loading of Lattice} A Bose Einstein condensate of typically $2 \times 10^5$ $^{87}$Rb-atoms (in the $F=2, m_F=2$ state) is produced in a magnetic trap with standard procedures at the crossing point of the interferometer branches. The magnetic trap is adiabatically deformed in order to prepare a nearly isotropic condensate fraction with $12\,\mu$m radius. Within 80 ms the lattice beam intensity is ramped up with a cubic spline to $\bar V_0/E_{\textrm{rec}} = 6.2$ with $\theta$ adjusted to $0.38\, \pi$. This leads to well depths of the $\mathcal{A}$-wells and $\mathcal{B}$-wells of $2.5\,E_{\textrm{rec}}$ and $11.4\,E_{\textrm{rec}}$, respectively. Excitation of the $P$-band is obtained by ramping $\theta$ within 0.2 ms to a value of $0.53\,\pi$, thus switching the $\mathcal{A}$-wells and $\mathcal{B}$-wells to $7.5\,E_{\textrm{rec}}$ and $5.0\,E_{\textrm{rec}}$, respectively.

\textit{Detection procedures} Momentum spectra are recorded by switching off the lattice beams in less than $1 \mu$s and subsequent free expansion of the atoms during 30 ms before an absorption image is recorded. For Brillouin zone mapping the lattice beam intensity is exponentially lowered with a time constant of $430\,\mu$s. Subsequently, ballistic expansion during 30 ms and absorption imaging is applied. The images shown in Figs. 1, 2 and 3 are averages over five individual measurements.

\emph{Acknowledgements} This work was partially supported by DFG (He2334/10-1) and Joachim Herz Stiftung.
We are grateful to C. Morais Smith, P. Schmelcher, L.-K. Lim, and C. Zimmermann for fruitful discussions.

\emph{Author contributions} All authors have equally contributed to this work.

\emph{Author Information} Correspondence and requests for materials should be addressed to A. H. (hemmerich@physnet.uni-hamburg.de).

\end{document}